\documentclass[preprint]{sigplanconf}
\usepackage{macros}

\begin{document}

\setlength{\pdfpageheight}{\paperheight}
\setlength{\pdfpagewidth}{\paperwidth}

\conferenceinfo{CONF 'yy}{Month d--d, 20yy, City, ST, Country} 
\copyrightyear{20yy} 
\copyrightdata{978-1-nnnn-nnnn-n/yy/mm} 
\doi{nnnnnnn.nnnnnnn}

% Uncomment one of the following two, if you are not going for the 
% traditional copyright transfer agreement.

%\exclusivelicense                % ACM gets exclusive license to publish, 
                                  % you retain copyright

%\permissiontopublish             % ACM gets nonexclusive license to publish
                                  % (paid open-access papers, 
                                  % short abstracts)

% \titlebanner{banner above paper title}        % These are ignored unless
% \preprintfooter{short description of paper}   % 'preprint' option specified.

\title{Symbolic Algorithms for Language Equivalence\\ and Kleene Algebra with Tests}
%\subtitle{}

%% anonymous submission
\authorinfo{Damien Pous%
  \thanks{We acknowledge support from the ANR projects 2010-BLAN-0305 PiCoq and 12IS02001 PACE.}}
           {CNRS, ENS de Lyon, UMR 5668, France}
           {Damien.Pous@ens-lyon.fr}

\maketitle

\begin{abstract}
  We first propose algorithms for checking language equivalence of
  finite automata over a large alphabet. We use symbolic automata,
  where the transition function is compactly represented using a
  (multi-terminal) binary decision diagrams (BDD). The key idea
  consists in computing a bisimulation by exploring reachable pairs
  symbolically, so as to avoid redundancies. This idea can be combined
  with already existing optimisations, and we show in particular a
  nice integration with the disjoint sets forest data-structure from
  Hopcroft and Karp's standard algorithm.

  Then we consider Kleene algebra with tests (KAT), an algebraic
  theory that can be used for verification in various domains ranging
  from compiler optimisation to network programming analysis.  This
  theory is decidable by reduction to language equivalence of automata
  on guarded strings, a particular kind of automata that have
  exponentially large alphabets. We propose several methods allowing
  to construct symbolic automata out of KAT expressions, based either
  on Brzozowski's derivatives or standard automata constructions.
  
  All in all, this results in efficient algorithms for deciding
  equivalence of KAT expressions.
\end{abstract}



\section{Introduction}
\label{sec:intro}

A wide range of algorithms in computer science build on the ability to
check language equivalence or inclusion of finite automata. In
model-checking for instance, one can build an automaton for a formula
and an automaton for a model, and then check that the latter is
included in the former. More advanced constructions need to build a
sequence of automata by applying a transducer, and to stop whenever
two subsequent automata recognise the same
language~\cite{BouajjaniHV04}. Another field of application is that of
various extensions of Kleene algebra, whose equational theories are
reducible to language equivalence of various kinds automata: regular
expressions and finite automata for plain Kleene
algebra~\cite{kozen94:ka:completeness}, ``closed'' automata for Kleene
algebra with converse~\cite{BES95,EB95}, or guarded string automata
for Kleene algebra with
tests (KAT)

The theory of KAT has been developed by Kozen et
al.~\cite{kozen:97:kat,cohenks96:kat:complexity,kozen08:kat:coalgebra},
it has received much attention for its applications in various
verification tasks ranging from compiler
optimisation~\cite{kozenp00:kat:compiler:opts} to program
schematology~\cite{angusk01:kat:schemato}, and very recently for
network programming analysis~\cite{NetKAT14,FKMST14a}. Like for Kleene
algebra, the equational theory of KAT is PSPACE-complete, making it a
challenging task to provide algorithms that are computationally
practical on as many inputs as possible. 

One difficulty with KAT is that the underlying automata work on an
input alphabet which is exponentially large in the number of variables
of the starting expressions. As such, it renders standard algorithms
for language equivalence intractable, even for reasonably small
inputs. This difficulty is shared with other fields where various
people proposed to work with \emph{symbolic automata} to cope with
large, or even infinite, alphabets~\cite{Bryant92,Veanes13}. By
symbolic automata, we mean finite automata whose transition function
is represented using a compact data-structure, typically binary
decision diagrams (BDDs)~\cite{Bryant86,Bryant92}, allowing the
explore the automata in a symbolic way.

D'Antoni and Veanes recently proposed a new minimisation algorithm for
symbolic automata~\cite{DAntoniV14}, which is much more efficient than
the adaptations of the traditional
algorithms~\cite{Moore56,Hopcroft:1971,Paige-Tarjan-PartitionRefinement}.
However, to our knowledge, the simpler problem of language equivalence
for symbolic automata has not been covered yet. We say `simpler'
because language equivalence can be reduced trivially to
minimisation---it suffices to minimise the automaton and to check
whether the considered states are equated, but minimisation has
complexity $n\mathrm{ln} n$ while Hopcroft and Karp's algorithm for
language equivalence~\cite{HopcroftKarp} is almost
linear~\cite{Tarjan75}.

Our main contributions are the following:
\begin{itemize}
\item We propose a simple coinductive algorithm for checking language
  equivalence of symbolic automata (Section~\ref{sec:symbolic}). This
  algorithm is generic enough to support various improvements that
  have been proposed in the literature for plain
  automata~\cite{CAV06,tacas10,DoyenR10,bp:popl13:hkc}.
\item We show how to combine binary decisions diagrams (BDD) and
  \emph{disjoint set forests}, the very elegant data-structure used by
  Hopcroft and Karp to defined their almost linear
  algorithm~\cite{HopcroftKarp,Tarjan75} for deterministic automata.
  This results in a new version of their algorithm, for symbolic
  automata (Section~\ref{ssec:dsf}).
\item We study several constructions for building efficiently a
  symbolic automaton out of a KAT expression (Section~\ref{sec:kat}):
  we consider a symbolic version of the extension of Brzozowski's
  derivatives~\cite{Brzozowski64} and Antimirov' partial
  derivatives~\cite{Antimirov96}, as well as a generalisation of Ilie
  and Yu's inductive construction~\cite{Ilie-yu-FollowAutomata}. The
  latter construction also requires us to generalise the standard
  procedure consisting in eliminating epsilon transitions.
\end{itemize}

\subsection*{Notation}

We denote sets by capital letters $X,Y,S,T\ldots$ and functions by
lower case letters $f,g,\dots$ Given sets $X$ and $Y$, $X\times Y$ is
their Cartesian product, $X\uplus Y$ is the disjoint union and $X^Y$
is the set of functions $f\colon Y\to X$. The collection of subsets of
$X$ is denoted by $\pow(X)$. For a set of letters $A$, $A^\star$
denotes the set of all finite words over $A$; $\epsilon$ the empty
word; and $w_1w_2$ the concatenation of words $w_1,w_2 \in
A^\star$. We use $2$ for the set $\set{0,1}$.

\section{Preliminary material}
\label{sec:prelim}

We first recall some standard definitions about finite automata and
binary decision diagrams. 

For finite automata, the only slight difference with the setting
described in~\cite{bp:popl13:hkc} is that we work with Moore
machines~\cite{Moore56} rather than automata: the accepting status of
a state is not necessarily a Boolean, but a value in a fixed yet
arbitrary set. Since this generalisation is harmless, we stick to the
standard automata terminology.

\subsection{Finite automata}
\label{ssec:finite:automata}

A deterministic finite automaton (DFA) over the input alphabet $A$ and
with outputs in $B$ is a triple $\tuple{S,t,o} $, where $S$ is a
finite set of states, $o\colon S \to B$ is the output function, and
$t\colon S \to S^A$ is the transition function which returns, for each
state $x$ and for each input letter $a\in A$, the next state
$t_a(x)$. For $a\in A$, we write $x\tr{a}x'$ for $t_a(x)=x'$.  For
$w\in A^\star$, we denote by $x\tr{w}x'$ for the least relation such
that (1) $x\tr{\epsilon}x$ and (2) $x\tr{aw'}x'$ if $x\tr{a}x''$ and
$x''\tr{w'}x'$.

The \emph{language} accepted by a state $x\in S$ of a DFA is the
function $\bb{x}\colon A^\star\to B$ defined as follows:
\begin{align*}
  \bb{x}(\epsilon) &= o(x)\enspace, &
  \bb{x}(a w) &= \bb{t_a(x)}(w)\enspace.
\end{align*}
(When the output set is $2$, these functions are indeed characteristic
functions of formal languages). Two states $x,y\in S$ are said to be
\emph{language equivalent} (written $x\sim y$) iff they accept the
same language.

\subsection{Coinduction}
\label{ssec:coinduction}

We then define bisimulations. We make explicit the underlying notion
of progression which we need in the sequel.
\begin{definition}[Progression, Bisimulation]\label{def:bisimulation}
  Given two relations $R,R'\subseteq S \times S$ on states, $R$
  \emph{progresses to} $R'$, denoted $R\prog R'$, if whenever
  $x\mathrel R y$ then
  \begin{enumerate}
  \item $o(x) = o(y) $ and
  \item for all $a\in A$, $t_a(x) \mathrel{R'} t_a(y)$. 
  \end{enumerate}
  A \emph{bisimulation} is a relation $R$ such that $R\prog R$.
\end{definition}

Bisimulation is a sound and complete proof technique for checking
language equivalence of DFA:

\begin{proposition}[Coinduction]
  \label{prop:bisimulation-langequivalence}
  Two states are language equivalent iff there exists a bisimulation
  that relates them.
\end{proposition}

Accordingly, we obtain the simple algorithm described in
Figure~\ref{alg:plain}, for checking language equivalence of two
states of a given automaton. (Note that to check language equivalence
of two states from two distinct automata, it suffices to consider the
disjoint union of the two automata.)

\begin{figure}
  \centering
\begin{ocaml}
type ('s,'b) dfa = {t: 's -> $A$ -> 's; o: 's -> 'b}

let equiv (M: ('s,'b) dfa) (x y: 's) =
    let r = Set.empty () in
    let todo = Queue.singleton (x,y) in
    while $\lnot$Queue.is_empty todo do 
        (* invariant: r \prog r \cup todo *) ?\label{line:plain:invariant}?
        let (x,y) = Queue.pop todo in
        if Set.mem r (x,y) then continue ?\label{line:plain:upto}?
        if M.o x <> M.o y then return false ?\label{line:plain:ce}?
        iter$_A$ (fun a -> Queue.push todo (M.t x a, M.t y a)) ?\label{line:plain:span}?
        Set.add r (x,y)
    done;
    return true
\end{ocaml}
  \caption{Simple algorithm for checking language equivalence.}
  \label{alg:plain}
\end{figure}

This algorithm works as follows: the variable \code{r} contains a
relation which is a bisimulation candidate and the variable
\code{todo} contains a queue of pairs that remain to be processed.  To
process a pair $(x,y)$, one first checks whether it already belongs to
the bisimulation candidate: in that case, the pair can be skipped
since it was already processed. Otherwise, one checks that the outputs
of the two states are the same ($o(x)=o(y)$), and one pushes all
derivatives of the pair to the \code{todo} queue: all pairs
$(t_a(x),t_a(y))$ for $a\in A$. The pair $(x,y)$ is finally added to
the bisimulation candidate, and we proceed with the remainder of the
queue.

The main invariant of the loop (line~\ref{line:plain:invariant}:
$\mathtt{r}\prog \mathtt{r}\cup \mathtt{todo}$) ensures that when
\code{todo} becomes empty, then \code{r} contains a bisimulation, and
the starting states were indeed bisimilar. Another invariant of the
loop is that for any pair $(x',y')$ in \code{todo}, there exists a
word $w$ such that $x\tr w x'$ and $y \tr w y'$. Therefore, if we
reach a pair of states whose outputs are
distinct---line~\ref{line:plain:ce}, then the word $w$ associated to
that pair witnesses the fact that the two initial states are not
equivalent.

\begin{remark}
  Note that such an algorithm can be modified to check for language
  inclusion in a straightforward manner: assuming an arbitrary
  preorder $\leq$ on the output set $B$, and letting language
  inclusion mean $x \leq y$ if for all $w\in A^\star$, $\bb
  x(w)\leq\bb y(w)$, it suffices to replace line~\ref{line:plain:ce}
  in Figure~\ref{alg:plain} by

  \smallskip
  {\rm \code{if $\lnot$(M.o x $\leq$ M.o y) then return false}.}
\end{remark}

\subsection{Up-to techniques}
\label{ssec:upto}

The previous algorithm can be enhanced by exploiting \emph{up-to
  techniques}~\cite{San98MSCS,pous:dsbook11}: an up-to technique is a
function $f$ on binary relations such that for any relation $R$ such
that $R\prog f(R)$ is contained in bisimilarity. Intuitively, such
relations, that are not necessarily bisimulations, are constrained
enough to be contained in bisimilarity.

Bonchi and Pous have recently shown~\cite{bp:popl13:hkc} that the
standard algorithm by Hopcroft and Karp~\cite{HopcroftKarp} actually
exploits such an up-to technique: on line~\ref{line:plain:upto},
rather than checking whether the processed pair is already in the
candidate relation \code{r}, Hopcroft and Karp check whether it
belongs to the equivalence closure of \code{r}. Indeed the function
$e$ mapping a relation to its equivalence closure is a valid up-to
technique, and this optimisation allows the algorithm to stop
earlier. Hopcroft and Karp moreover use an efficient data-structure to
perform this check in almost constant time~\cite{Tarjan75}:
\emph{disjoint sets forests}. We recall this data-structure in
Section~\ref{ssec:dsf}.

Other examples of valid up-to techniques include context-closure, as
used in antichain based algorithms~\cite{CAV06,tacas10,DoyenR10}, or
congruence closure~\cite{bp:popl13:hkc}, which combines both
context-closure and equivalence closure. These techniques however
require to work with automata whose state carry a semi-lattice
structure, as is typically the case for a DFA obtained from a
non-deterministic automaton, through the powerset construction.

\subsection{Binary decision diagrams}
\label{ssec:bdd}

Assume an ordered set $(A,<)$ and an arbitrary set $B$. Binary
decision diagrams are directed acyclic graphs that can be used to
represent functions of type $2^A\to B$. When $B=2$ is the two
elements set, BDDs thus intuitively represent Boolean formulas with
variables in $A$.

Formally, a \emph{(multi-terminal, ordered) binary decision diagram}
(BDD) is a pair $(N,c)$ where $N$ is a finite set of nodes and $c$ is
a function of type $N \to B \uplus A\times N\times N$ such that if
$c(n)=(a,l,r)$ and either $c(l)=(a',\_,\_)$ or $c(r)=(a',\_,\_)$, then
$a<a'$.

The condition on $c$ ensures that the underlying graph is acyclic,
which make it possible to associate a function $\sem n\colon 2^A \to
B$ to each node $n$ of a BDD:
\begin{align*}
  \sem n(\alpha) &= 
  \begin{cases}
    b & \text{if }c(n)=b\in B\\
    \sem l(\alpha) &\text{if }c(n)=(a,l,r) \text{
      and } \alpha(a)=0\\
    \sem r(\alpha) &\text{if }c(n)=(a,l,r) \text{
      and } \alpha(a)=1
  \end{cases}
\end{align*}

Let us now recall the standard graphical representation of BDDs:
\begin{itemize}
\item A node $n$ such that $c(n)=b\in B$ is represented by a square
  box labelled by $b$.
\item A node $n$ such that $c(n)=(a,l,r)\in A\times N\times N$ is a
  decision node, which we picture by a circle labelled by $a$, with
  a dashed arrow towards the \emph{left child} $(l)$ and a plain arrow
  towards the \emph{right child} $(r)$.
\end{itemize}
For instance, the following drawing represents a BDD with three nodes;
its top-most node denotes the function given on the right-hand side.
\begin{align*}
  \begin{minipage}{.3\linewidth}
    \includegraphics[width=2cm]{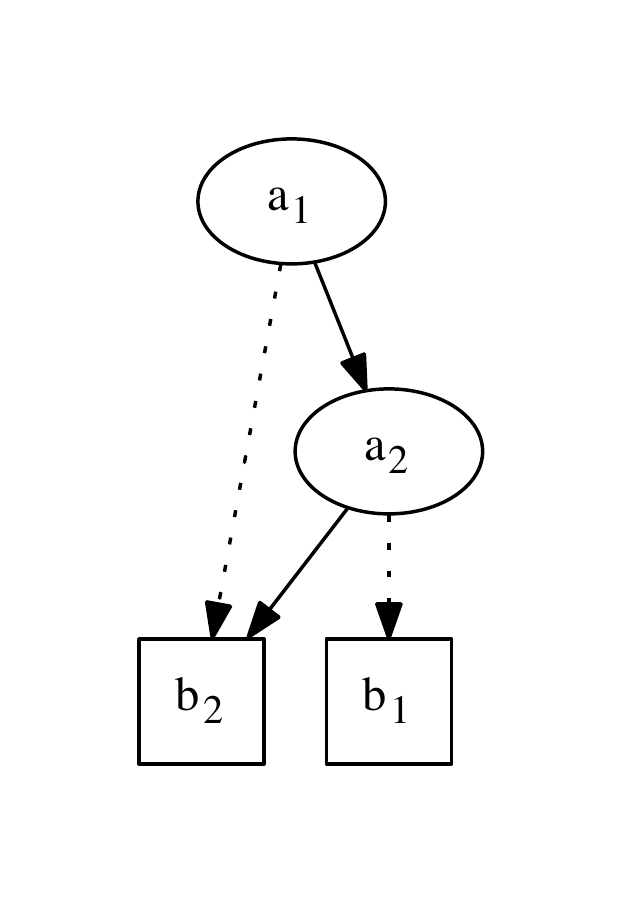}
  \end{minipage}
  && \alpha\mapsto
  \begin{cases}
    b_1 &\text{if }\alpha(a_1)=1 \text{ and } \alpha(a_2)=0\\
    b_2 &\text{otherwise}
  \end{cases}
\end{align*}

A BDD is \emph{reduced} if $c$ is injective, and $c(n)=(a,l,r)$
entails $l\neq r$. (The above example BDD is reduced.) Any BDD can be
transformed into a reduced one. When $A$ is finite, reduced (ordered)
BDD nodes are in one-to-one correspondence with functions from $2^A$
to $B$~\cite{Bryant86,Bryant92}. The main interest in this
data-structure is that it is often extremely compact.

In the sequel, we only work with reduced ordered BDDs, which we simply
call BDDs. We denote by $\bdd A B$ the set of nodes of a large enough
BDD with values in $B$, and we let $\mes f$ denote the unique BDD node
representing a given function $f\colon 2^A\to B$. This notation is
useful to give abstract specifications to BDD operations: in the
sequel, all usages of this notation actually underpin efficient BDD
operations.

\paragraph{Implementation.}
To better explain parts of the proposed algorithms, we give a simple
implementation of BDDs in Figure~\ref{alg:bdd}.

\begin{figure}
\begin{ocaml}
type 'b node = 'b descr hash_consed
and  'b descr = V of 'b | N of $A$ ** 'b node ** 'b node

val hashcons: 'b descr -> 'b node
val c: 'b node -> 'b descr
val memo_rec: (('a -> 'b -> 'c) -> 'a -> 'b -> 'c) -> 'a -> 'b -> 'c

let constant v = hashcons (V v)
let node a l r = if l==r then l else hashcons (N(a,l,r))

let apply (f: 'a -> 'b -> 'c): 'a node -> 'b node -> 'c node =
  memo_rec (fun app x y -> 
  match c(x), c(y) with
    | V v, V w -> constant (f v w)
    | N(a,l,r), V _ -> node a (app l y) (app r y)
    | V _, N(a,l,r) -> node a (app x l) (app x r)
    | N(a,l,r), N(a',l',r') -> 
        if a=a' then node a  (app l l') (app r r')
        if a<a' then node a  (app l y ) (app r y )
        if a>a' then node a' (app x l') (app x r'))
\end{ocaml}
  \caption{An implementation of BDDs.}
  \label{alg:bdd}
\end{figure}

The type for BDD nodes is given first: we use Filli\^atre's
hash-consing library~\cite{FilliatreC06} to enforce unique
representation of each node, whence the two type declarations and the
two conversion functions \code{hashcons} and \code{c} between those
types. The third utility function \code{memo_rec} is just a convenient
operator for defining recursive memoised functions.

The function \code{constant} creates a constant node, making sure it
was not already created. The function \code{node} creates a new
decision node, unless that node is useless and can be replaced by one
of its two children. %This ensures that created nodes are reduced.
The generic function \code{apply} is central to
BDDs~\cite{Bryant86,Bryant92}: many operations are just instances of
this function. Its specification is the following:
\begin{align*}
  \mathtt{apply}~f~x~y &= \mes{\alpha\mapsto f(\sem x(\alpha))(\sem y(\alpha))}
\end{align*}
This function is obtained by ``zipping'' the two BDDs together until a
constant is reached. Memoisation is used to exploit sharing and to
avoid performing the same computations again and again.

\medskip

Suppose now that we want to define logical disjunction on Boolean BDD
nodes. Its specification is the following: 
\begin{align*}
  x \lor y &= \mes{\alpha\mapsto \sem n(\alpha)\lor \sem m(\alpha)}.
\end{align*}
We can thus simply use the \code{apply} function, applied to the
Boolean disjunction function:
\begin{ocaml}
let dsj: bool node -> bool node -> bool node = apply (||)  
\end{ocaml}
Note that this definition could actually be slightly optimised by
inlining \code{apply}'s code, and noticing that the result is already
known whenever one of the two arguments is a constant:
\begin{ocaml}
let dsj: bool node -> bool node -> bool node =
  memo_rec (fun dsj x y -> 
  match c(x), c(y) with
    | V true, _ | _, V false -> x
    | _, V true | V false, _ -> y
    | N(a,l,r), N(a',l',r') -> 
        if a=a' then node a  (dsj l l') (dsj r r')
        if a<a' then node a  (dsj l y ) (dsj r y )
        if a>a' then node a' (dsj x l') (dsj x r'))
\end{ocaml}
We ignore such optimisations in the sequel, for the sake of clarity.

\section{Symbolic automata}
\label{sec:symbolic}

A standard
technique~\cite{Bryant92,HenriksenJJKPRS95,Veanes13,DAntoniV14} for
working automata over a large input alphabet consists in using BBDs to
represent the transition function: a \emph{symbolic DFA} with output
set $B$ and input alphabet $A'=2^A$ for some set $A$ is a triple
$\tuple{S,t,o}$ where $S$ is the set of states, $t\colon S\to \bdd A
S$ maps states into nodes of a BDD over $A$ with values in $S$, and
$o\colon S \to B$ is the output function.

\begin{figure}
  \centering
  \includegraphics[width=7cm]{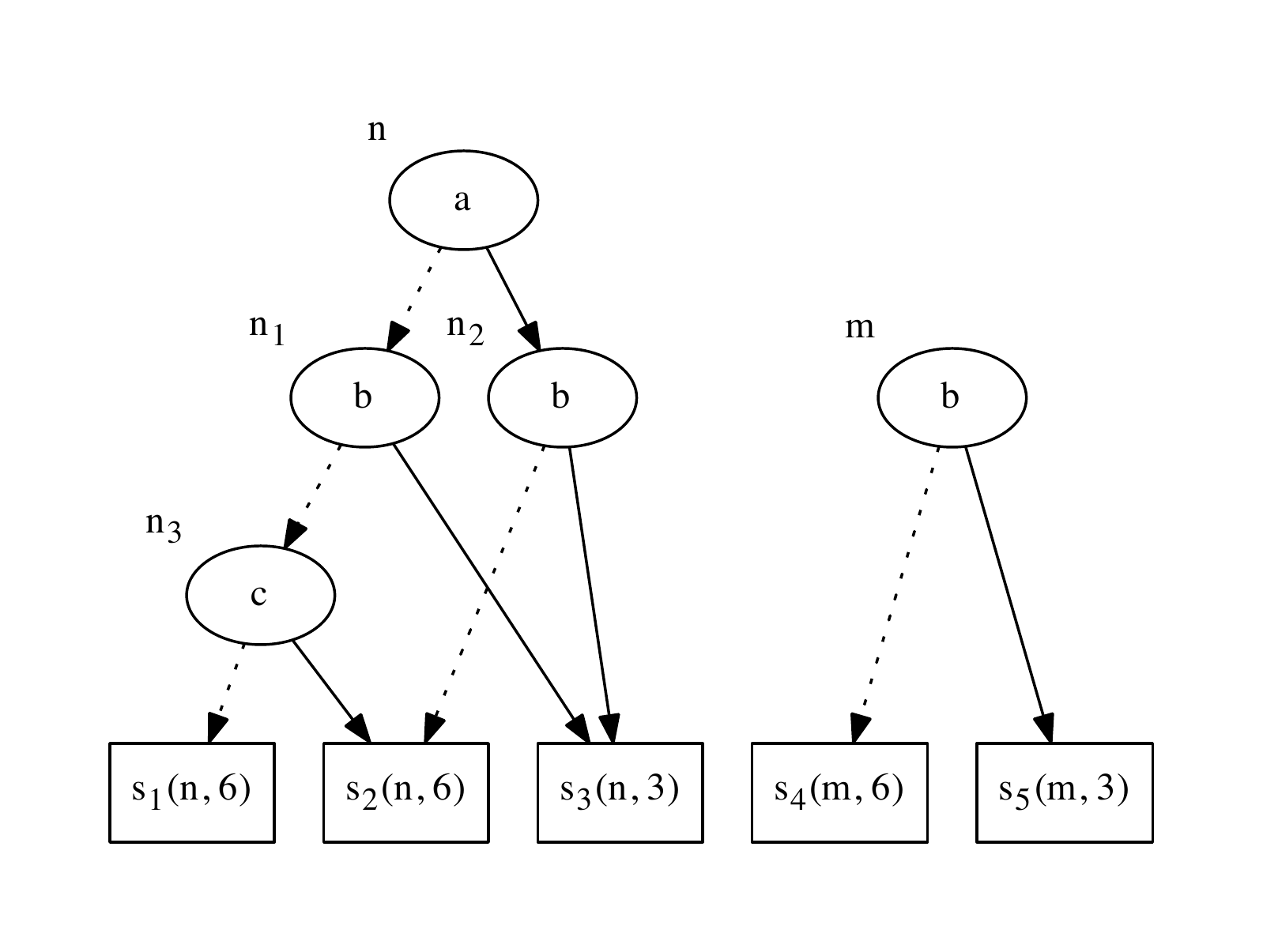}
  \begin{align*}
    \begin{array}{c|l@{\,}l@{\,}l@{\,}l@{\,}l@{\,}l@{\,}l@{\,}l|l@{\,}l@{\,}l@{\,}l@{\,}l@{\,}l@{\,}l@{\,}l}
      &\multicolumn{8}{c}{s_1, s_2, s_3}&\multicolumn{8}{c}{s_4, s_5}\\\hline
      a&0&0&0&0&1&1&1&1&0&0&0&0&1&1&1&1\\
      b&0&0&1&1&0&0&1&1&0&0&1&1&0&0&1&1\\
      c&0&1&0&1&0&1&0&1&0&1&0&1&0&1&0&1\\\hline
      t&
      s_1&s_2&s_3&s_3&s_2&s_2&s_3&s_3&
      s_4&s_4&s_5&s_5&s_4&s_4&s_5&s_5\\
    \end{array}
  \end{align*}
  \label{fig:symbolic:dfa}
  \caption{A symbolic DFA with five states.}
\end{figure}

Such a symbolic DFA is depicted in Figure~\ref{fig:symbolic:dfa}. It
has five states, input alphabet $2^{\set{a,b,c}}$, and natural numbers
as output set.  We represent the BDD graphically; for each state, we
write the values of $t$ and $o$ together with the name of the state,
in the corresponding square box. The explicit transition table is
given below the drawing.

The simple algorithm described in Figure~\ref{alg:plain} is not
optimal when working with such symbolic DFAs: at each non-trivial
iteration of the main loop, one goes through all letters of $A'=2^A$
to push all the derivatives of the current pair of states to the queue
\code{todo} (line~\ref{line:plain:span}), resulting in a lot of
redundancies. 

Suppose for instance that we run the algorithm on the DFA of
Figure~\ref{fig:symbolic:dfa}, starting from states $s_1$ and
$s_4$. After the first iteration, \code{r} contains the pair
$(s_1,s_4)$, and the queue \code{todo} contains eight pairs: 
\begin{align*}
  (s_1,s_4), (s_2,s_4), (s_3,s_5), (s_3,s_5), (s_2,s_4), (s_2,s_4), (s_3,s_5), (s_3,s_5)
\end{align*}
Assume that elements of this queue are popped from left to right.  The
first two elements are removed during the next two iterations, since
$(s_1,s_4)$ already is in \code{r}. Then $(s_2,s_4)$ is processed: it
is added to \code{r}, and the above eight pairs are appended again to
the queue, which now has thirteen elements. The following pair is
processed similarly, resulting in a queue with twenty ($13-1+8$)
pairs. Since all pairs of this queue are already in \code{r}, it is
finally emptied through twenty iterations, and the algorithm returns
true.

Note that it would be even worse if the input alphabet was actually
declared to be $2^{\set{a,b,c,d}}$: even though the bit $d$ of all
letters is irrelevant for the considered DFA, each non-trivial
iteration of the algorithm would push even more copies of each pair to
the \code{todo} queue.

What we propose here is to exploit the symbolic representation, so
that a given pair is pushed only once. Intuitively, we want to
recognise that starting from the pair of nodes $(n,m)$, the letters
$010$, $011$, $110$ and $111$ are equivalent\footnote{Letters being
  elements of $2^{\set{a,b,c}}$ here, we represent them with
  bit-vectors of length three}, since they yield to the same pair,
$(s_3,s_5)$. Similarly, the letters $001$, $100$, and $101$ are
equivalent: they yield to the pair $(s_2,s_4)$.

This idea is easy to implement using BDDs: like for the \code{apply}
function (Figure~\ref{alg:bdd}), it suffices to zip the two BBDs
together, and to push pairs when we reach two leaves. We use for that 
the procedure \code{pairs} from Figure~\ref{alg:pairs}, which
successively applies a given function to all pairs reachable from two
nodes. Its code is almost identical to \code{apply}, except that
nothing is constructed (and memoisation is just used to remember those
pairs that have already been visited).

\begin{figure}
  \centering
\begin{ocaml}
let pairs (f: 'a ** 'b -> unit): 'a node -> 'b node -> unit =
   memo_rec (fun pairs x y -> 
   match c(x), c(y) with
      | V v, V w -> f (v,w)
      | V _, N(_,l,r) -> pairs x l; pairs x r
      | N(_,l,r), V _ -> pairs l y; pairs r y
      | N(a,l,r), N(a',l',r') -> 
          if a=a' then pairs l l'; pairs r r' 
          if a<a' then pairs l y ; pairs r y
          if a>a' then pairs x l'; pairs x r') 
\end{ocaml}
  \label{alg:pairs}
  \caption{Iterating over the set of pairs reachable from two nodes.}
\end{figure}

We finally modify the simple algorithm from
Section~\ref{ssec:finite:automata} by using this procedure on
line~\ref{line:symb:span}: we obtain the code given in
Figure~\ref{alg:symb:equiv}.
\begin{figure}
  \centering
\begin{ocaml}
type ('s,'b) sdfa = {t: 's -> 's bdd; o: 's -> 'b}

let symb_equiv (M: ('s,'b) sdfa) (x y: 's) =
    let r = Set.empty () in
    let todo = Queue.singleton (x,y) in
    let push_pairs = pairs (Queue.push todo) in ?\label{line:more:memo}?
    while $\lnot$Queue.is_empty todo do 
        let (x,y) = Queue.pop todo in
        if Set.mem r (x,y) then continue
        if M.o x <> M.o y then return false ?\label{line:symb:ce}?
        push_pairs (M.t x) (M.t y) ?\label{line:symb:span}?
        Set.add r (x,y)
    done;
    return true
\end{ocaml}
  \caption{Symbolic algorithm for checking language equivalence.}
  \label{alg:symb:equiv}
\end{figure}
We apply \code{pairs} to its first argument once and for all
(line~\ref{line:more:memo}), so that we maximise memoisation: a pair
of nodes that has been visited in the past will never be visited
again, since all pairs of states reachable from that pair of nodes is
already guaranteed to be processed. (As an invariant, we have that all
pairs reachable from a pair of nodes memoised in \code{push_pairs}
appear in \code{r \cup todo}.)

Let us illustrate this algorithm by running it on the DFA from
Figure~\ref{fig:symbolic:dfa}, starting from states $s_1$ and $s_4$ as
previously. During the first iteration, the pair $(s_1,s_4)$ is added
to \code{r}, and \code{push_pairs} is called on the pair of nodes
$(n,m)$. This call virtually results in building the following BDD,
\begin{center}
  \includegraphics[width=3.8cm]{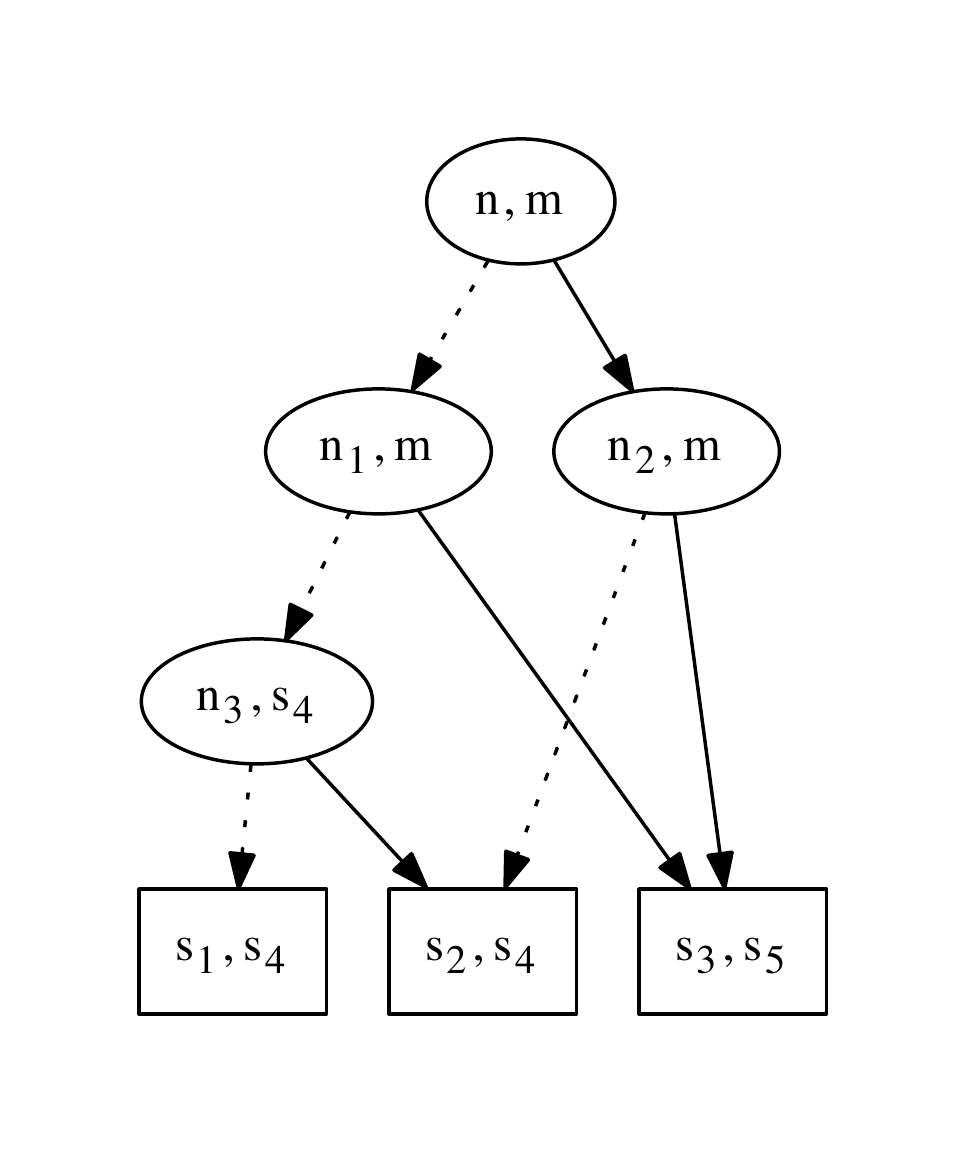}
\end{center}
so that the following three pairs are pushed to \code{todo}.
\begin{align*}
  (s_1,s_4), (s_2,s_4), (s_3,s_5)
\end{align*}
The first pair is removed by a trivial iteration: $(s_1,s_4)$ already
belongs to \code{r}. The two other pairs are processed by adding them to
\texttt{r}, but without pushing any new pair to \code{todo}: thanks to
memoisation, the two expected calls to \code{push_pairs n m} are
skipped.

All in all, each reachable pair is pushed only once to the \code{todo}
queue. More importantly, the derivatives of a given pair are explored
symbolically. In particular, the algorithm would execute exactly in
the same way, even if the alphabet was actually declared to be much
larger (for instance because the considered states were part of a
bigger automaton with more letters).

\subsection{Displaying symbolic counter-examples.}
\label{ssec:ce}

Another advantage of this new algorithm is that it can easily be
instrumented to produce concise counter-examples in case of failure.
Consider for instance the following automaton
\begin{center}
  \includegraphics[width=1.1\linewidth]{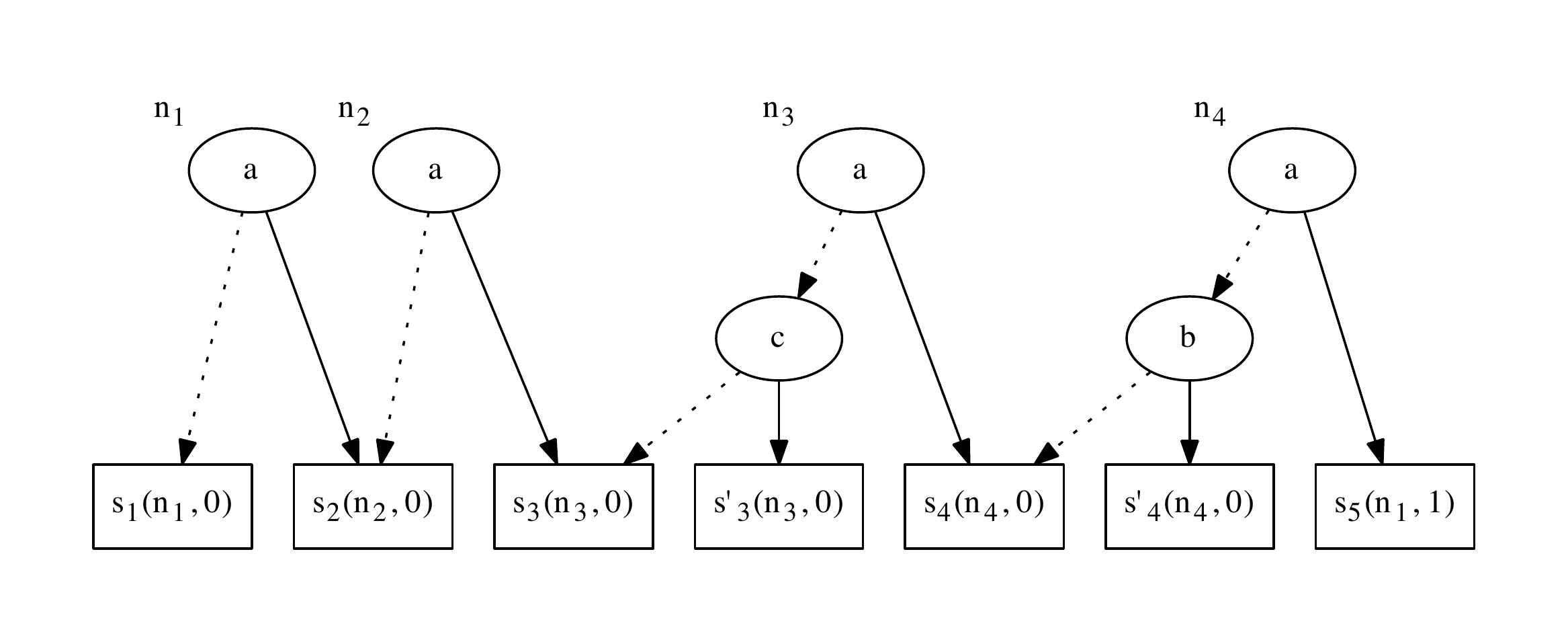}
\end{center}
Intuitively, the states $s_1$ and $s_2$ are not equivalent because
$s_2$ can take three transitions to reach $s_5$, with output $1$,
while $s_1$ cannot reach $s_5$ in three transitions.

More formally, the word $100~100~100$ over $2^{\set{a,b,c}}$ is a 
counter-example: we have 
\begin{align*}
  \bb{s_1}(100~100~100)&=\bb{s_2}(100~100)=\bb{s_3}(100)=o(s_4)=0\\
  \bb{s_2}(100~100~100)&=\bb{s_3}(100~100)=\bb{s_4}(100)=o(s_5)=1
\end{align*}
But there are plenty of other counter-examples of length three: it
suffices that $a$ be assigned true in the three letters, the value of
the bits $b$ and $c$ does not change the above computation. As a
consequence, this counter-example is best described as the word $a\,
a\, a$, whose letters are Boolean formulas in conjunctive normal form
indicating the least requirements to get a counter example.

The algorithm from Figure~\ref{alg:symb:equiv} makes it possible to
give this information back to the user: 
\begin{itemize}
\item modify the queue \emph{todo} to store triples $(w,x,y)$ where $(x,y)$
  is a pair of states to process, and $w$ is the associated potential
  counter-example;
\item modify the function \code{pairs} (Figure~\ref{alg:pairs}), so
  that it uses an additional argument to record the encountered node
  labels, with negative polarity when going through the recursive call
  for the left children, and positive polarity for the right children;
\item modify line~\ref{line:plain:ce} of the main algorithm to return
  the symbolic word associated current pair when the output test
  fails.
\end{itemize}

\subsection{Non-deterministic automata}
\label{ssec:det}

Standard coinductive algorithms for DFA can be applied to
non-deterministic automata (NFA) by using the \emph{powerset
  construction}, on the fly. This construction transforms a
non-deterministic automaton into a deterministic one; we extend it to
symbolic automata in the straightforward way.

A \emph{symbolic NFA} is a tuple $\tuple{S,t,o}$ where $S$ is the set
of states, $o\colon S\to B$ is the output function, and $t\colon S\to
\bdd A{\pow S}$ maps a state and a letter of the alphabet $A'=2^A$ to
a set of possible successor states, using a symbolic representation.

Assuming such an NFA, one defines a symbolic DFA
$\tuple{\pow S,t^\sharp,o^\sharp}$ as follows:
\begin{align*}
  t^\sharp(\set{x_1,\dots,x_n})&\eqdef t(x_1)\bddcup\dots\bddcup t(x_n) \\
  o^\sharp(\set{x_1,\dots,x_n})&\eqdef o(x_1)\bddlor\dots\vee o(x_n)
\end{align*}
(Where $\bddcup$ denotes the pointwise union of two BDDs over sets:
$n\bddcup m = \mes{\phi\mapsto\sem n(\phi)\cup \sem m(\phi)}$.)

\subsection{Hopcroft and Karp: disjoint sets forests}
\label{ssec:dsf}

The previous algorithm can be freely enhanced by using up-to
techniques, as described in Section~\ref{ssec:upto}: it suffices to
modify line~\ref{line:plain:upto} to skip pairs more or less
aggressively, according to the chosen up-to technique.

The up-to-equivalence technique used in Hopcroft and Karp's algorithm
can however be integrated in a deeper way, by exploiting the fact that
we work with BDDs. This leads to a second algorithm, which we describe
in this section.

Let us first recall \emph{disjoint sets forests}, the data structure
used by Hopcroft and Karp to represent equivalence classes. This
standard data-structure makes it possible to check whether two
elements belong to the same class and to merge two equivalence
classes, both in almost constant amortised time~\cite{Tarjan75}.

The idea consists in storing a partial map from elements to elements
and whose underlying graph is acyclic. An element for which the map is
not defined is the \emph{representative} of its equivalence class, and
the representative of an element pointing in the map to some $y$ is the
representative of $y$. Two elements are equivalent if and only if they
lead to the same representative, and to merge two equivalence classes,
it suffices to add a link from the representative of one class to the
representative of the other class. Two optimisations are required to
obtain the announced theoretical complexity:
\begin{itemize}
\item when following the path leading from an element to its
  representative, one should compress it in some way, by modifying the
  map so that the elements in this path become closer to their
  representative. There are various ways of compressing paths, in the
  sequel, we use the method called \emph{halving}~\cite{Tarjan75};
\item when merging two classes, one should make the smallest one point
  to the biggest one, to avoid generating too many long paths. Again,
  there are several possible heuristics, but we elude this point in
  the sequel.
\end{itemize}

As explained above, the simplest thing to do would be to replace the
bisimulation candidate $r$ from Figure~\ref{alg:symb:equiv} by a
disjoint sets forest over the states of the considered automaton.

The new idea consists in relating the BBD nodes of the symbolic
automaton rather that just its states (i.e., just the BDD leaves).  By
doing so, one avoids visiting pairs of nodes that have already been
visited up to equivalence.

Concerning the implementation, we first introduce a variant of the
function \code{pair} in Figure~\ref{alg:dsf:pairs}, which uses
disjoint sets forest rather than plain memoisation. 
\begin{figure}
  \centering
\begin{ocaml}
let pairs' (f: 'b ** 'b -> unit): 'b node -> 'b node -> unit =
  (* the disjoint sets forest *)
  let m = Hmap.empty() in       
  let link x y = Hmap.add m x y in
  (* representative of a node *)
  let rec repr x =              
    match Hmap.get m x with
      | None -> x
      | Some y -> match Hmap.get m y with
          | None -> y
          | Some z -> link x z; repr z
  in 
  let rec pairs x y = 
    let x = repr x in
    let y = repr y in
    if x <> y then
    match c(x), c(y) with
      | V v, V w -> link x y; f (v,w) ?\label{line:freelink1}?
      | V _, N(_,l,r) -> link y x; pairs x l; pairs x r
      | N(_,l,r), V _ -> link x y; pairs l y; pairs r y
      | N(a,l,r), N(a',l',r') -> 
          if a=a' then link x y; pairs l l'; pairs r r' ?\label{line:freelink2}?
          if a<a' then link x y; pairs l y ; pairs r y
          if a>a' then link y x; pairs x l'; pairs x r')    
  in pairs
\end{ocaml}
  \label{alg:dsf:pairs}
  \caption{Iterating over the set of pairs reachable from two nodes,
    optimised using disjoint set forests.}
\end{figure}
This function first creates an empty forest (we use for that use
Filli\^atre's implementation of maps over hash-consed values). The
function \code{link} adds a link between two representatives; the
recursive terminal function \code{repr} looks for the representative
of a node and implements halving. The function \code{pairs'} is
defined similarly as \code{pairs}, except that it first takes the
representative of the two given nodes, and that it adds a link from
one to the other before recursing.

Those links can be put in any direction on lines~\ref{line:freelink1}
and~\ref{line:freelink2}, and we should actually use an appropriate
heuristic to take this decision, as explained above. In the four other
cases, we put a link either from the node to the leaf, or from the
node with the smallest label to the node with the biggest label. By
proceeding this way, we somehow optimise the BDD, by leaving as few
decision nodes as possible.

It is however important to notice that there is actually no choice
left in those four cases: we work implicitly with the optimised BDD
obtained by mapping all nodes to their representatives, so that we
have to maintain the invariant that this optimised BDD is ordered and
acyclic. (Notice that on the contrary, this optimised BDD need not be
reduced anymore: the children of given a node might be silently
equated, and a node might have several representations since its
children might be silently equated with the children of another node
with the same label)

We finally obtain the algorithm given in Figure~\ref{alg:dsf:equiv}.
\begin{figure}
  \centering
\begin{ocaml}
let dsf_equiv (M: ('s,'b) sdfa) (x y: 's) =
  let todo = Queue.singleton (x,y) in
  let push_pairs = pairs' (Queue.push todo) in
  while $\lnot$Queue.is_empty todo do 
    let (x,y) = Queue.pop todo in
    if M.o x <> M.o y then return false
    push_pairs (M.t x) (M.t y) 
  done;
  return true
\end{ocaml}
  \caption{Symbolic algorithm optimised with disjoint set forests.}
  \label{alg:dsf:equiv}
\end{figure}
It is similar to the previous one (Figure~\ref{alg:symb:equiv}),
except that we use the above new function \code{pairs'} to push pairs
into the \code{todo} queue, and that we no longer need to store the
bisimulation candidate \code{r}: this relation is subsumed by the
restriction of the disjoint set forests to BDD leaves.

\medskip

If we execute this algorithm on the symbolic DFA from
Figure~\ref{fig:symbolic:dfa}, between states $s_1$ and $s_4$, we
obtain the disjoint set forest depicted below using dashed red
arrows. This is actually corresponds to the pairs which would be
visited by the first symbolic algorithm (Figure~\ref{alg:symb:equiv}).
\begin{center}
  \includegraphics[width=7cm]{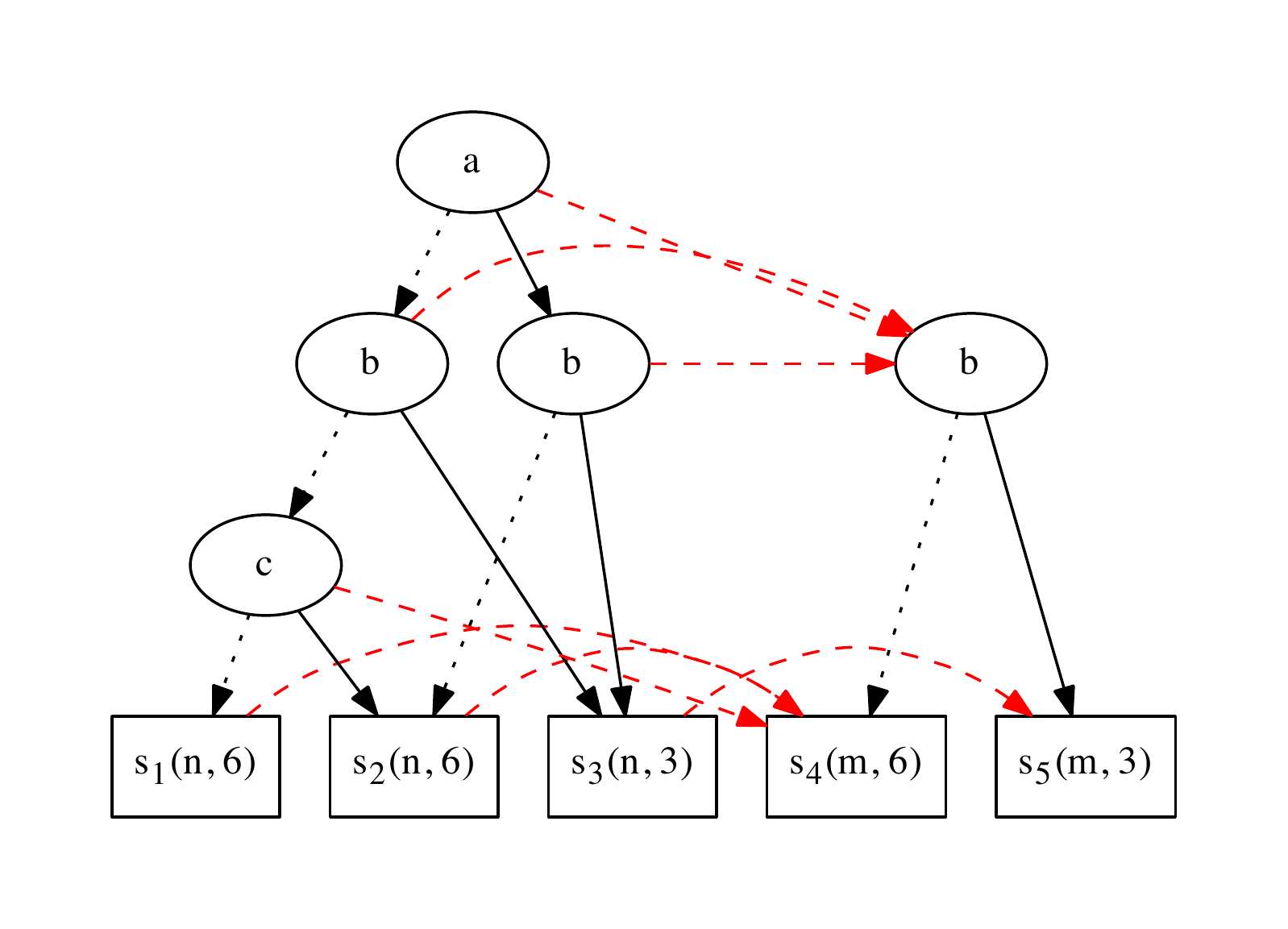}
\end{center}

If instead we start from nodes $n1$ and $m1$ in the following partly
described automaton, we would get the disjoint set forest depicted
similarly in red, while the first algorithm would go through all blue
pairs, one of which contains is superfluous.
\begin{center}
  \includegraphics[width=7cm]{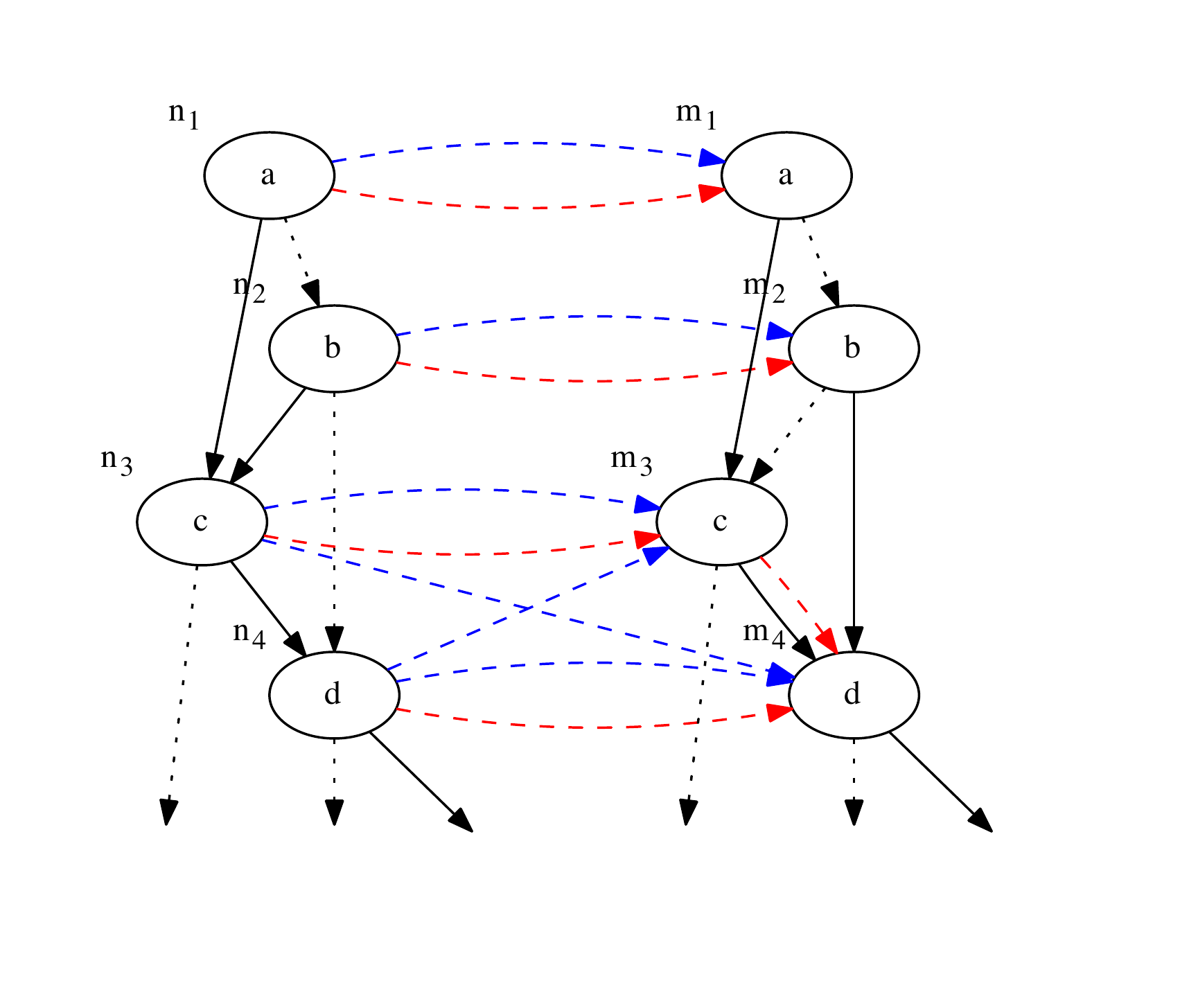}
\end{center}

\section{Kleene algebra with tests}
\label{sec:kat}

Now we consider Kleene algebra with tests, for which we provide
several automata constructions that allow one to use the previous
symbolic algorithms.

\medskip

A \emph{Kleene algebra with tests} (KAT) is a tuple 
$\tuple{X,B,{\cdot},{+},{\cdot^\star},\lnot,1,0}$ such that
\begin{enumerate}[(i)]
\item $\tuple{X,{\cdot},{+},{\cdot^\star},1,0}$ is a Kleene
  algebra~\cite{kozen94:ka:completeness}, i.e., an idempotent semiring
  with a unary operation, called ``Kleene star'', satisfying the
  following axiom and inference rules:
 \begin{mathpar}
    1+x\cdot{}x^\star \le x^\star \and
    % 1+x^\star\cdot{}x \le x^\star  \and
    \inferrule{y\cdot{}x \le x}{y^\star\cdot{}x \le x} \and %
    \inferrule{x\cdot{}y \le x}{x\cdot{}y^\star \le x}
  \end{mathpar}
  (The preorder $(\le)$ being defined by $x\le y ~ \eqdef ~ x+y = y$.)
\item $B\subseteq X$
\item $\tuple{B,{\cdot},{+},{\neg},1,0}$ is a Boolean algebra.
\end{enumerate}

The elements of the set $B$ are called ``tests''; we denote them by
$\phi,\psi$. The elements of $X$, called ``Kleene elements'', are
denoted by $x,y,z$. We sometimes omit the operator ``$\cdot$'' from
expressions, writing $xy$ for $x\cdot y$. The following (in)equations
illustrate the kind of laws that hold in all Kleene algebra with tests:
\begin{mathpar}{}
  \phi+\neg \phi = 1 \and %
  \phi\cdot(\neg \phi + \psi) = \phi\cdot \psi =
  \neg(\neg \phi + \neg \psi) \\
  x^\star x^\star = x^\star \and %
  (x+y)^\star = x^\star(yx^\star)^\star \and
  (x+x x y)^\star \le (x+x y)^\star\\  
  \phi\cdot(\neg \phi \cdot x)^\star = \phi \and %
  \phi\cdot(\phi\cdot x \cdot \neg \phi+\neg \phi\cdot y \cdot \phi)^\star\cdot\phi
  \le (x\cdot y)^\star %
\end{mathpar}
The laws from the first line come from the Boolean algebra structure,
while the ones from the second line come from the Kleene algebra
structure. The two laws from the last line require both Boolean
algebra and Kleene algebra reasoning.

\paragraph{Binary relations.} 
Binary relations form a Kleene algebra with tests; this is the main
model we are interested in, in practice. The Kleene elements are the
binary relations over a given set $S$, the tests are the predicates
over this set, encoded as sub-identity relations, and the star of a
relation is its reflexive transitive closure.

This relational model is typically used to interpret imperative
programs: such programs are state transformers, i.e., binary relations
between states, and the conditions used to define the control-flow of
these programs are just predicates on states. Typically, a program
``\code{while $\phi$ do p}'' is interpreted through the KAT expression
$(\phi\cdot p)^\star\cdot\lnot\phi$.

\paragraph{KAT expressions.}

We denote by $Rel(V)$ the set of \emph{regular expressions} over a set
$V$:
\begin{align*}
  x,y ::= v\in V \OR x+y \OR x\cdot y \OR x^\star\enspace.
\end{align*}

Assuming a set $A$ of elementary tests, we denote by $B(A)$ the set of
\emph{Boolean expressions} over $A$:
\begin{align*}
  \phi,\psi ::= a\in A \OR 1 \OR 0 
  \OR \phi\land\phi \OR \phi\lor\phi \OR \lnot\phi
\end{align*}

Further assuming a set $\Sigma$ of letters (or atomic Kleene
elements), a \emph{KAT expression} is a regular expression over the
disjoint union $\Sigma\uplus B(A)$. Note that the constants $0$ and
$1$ from the signature of KAT, and usually found in the syntax of
regular expressions, are represented here by injecting the
corresponding tests.

\paragraph{Guarded string languages.}
Guarded string languages are the natural generalisation of string
languages for Kleene algebra with tests. We briefly define them.

An \emph{atom} is a valuation from elementary tests to Booleans; it
indicates which of these tests are satisfied. We let $\alpha,\beta$
range over atoms, the set of which is denoted by $At$: $At=2^A$.
A Boolean formula $\phi$ is \emph{valid} under an atom $\alpha$,
denoted by $\alpha\vDash\phi$, if $\phi$ evaluates to true under the
valuation $\alpha$.

A \emph{guarded string} %~\cite{Kaplan69}:
is an alternating sequences of atoms and letters, both starting and
ending with an atom:
\begin{align*}
  \alpha_1,p_1,\alpha_2,\dots,\alpha_n,p_n,\alpha_{n+1}\enspace.
\end{align*}

The concatenation $u\ast v$ of two guarded strings $u,v$ is a partial
operation: it is defined only if the last atom of $u$ is equal to the
first atom of $v$; it consists in concatenating the two sequences and
removing one copy of the shared atom in the middle.

To any KAT expression, one associates a \emph{guarded string
  language}, i.e., a set of guarded strings, as follows:
\begin{align*}
  G(\phi) &= \set{\alpha\in At \mid \alpha \vDash \phi} 
  \tag*{$(\phi\in B(A))$}
  \\
  G(p) &= \set{\alpha p \beta \mid \alpha,\beta\in At} 
  \tag*{$(p\in \Sigma)$}
  \\
  G(x+y) &= G(x) \cup G(y) \\
  G(xy) &= \set{u\ast v \mid u\in G(x), v\in G(y)} \\
  G(x^\star) &= \set{u_1\ast \dots \ast u_n \mid \exists u_1\dots u_n, \forall i\leq n, u_i\in G(x)}
\end{align*}

\paragraph{KAT Completeness.}

Kozen and Smith proved that the equational theory of Kleene algebra
with tests is complete over the relational
model~\cite{kozens96:kat:completeness:decidability}: any equation that
holds universally in this model can be proved from the axioms of
KAT. Moreover, two expressions are provably equal if and only if they
denote the same language of guarded strings. By a simple reduction to
automata theory this gives algorithms to decide the equational theory
of KAT. Now we study several such algorithms, and we show each time
how to exploit symbolic representations to make them efficient.

\subsection{Brzozowski's derivatives}
\label{ssec:brz}

Derivatives were introduced by Brzozowski~\cite{Brzozowski64} for
(plain) regular expressions; they make it possible to define a
deterministic automaton where the states of the automaton are the
regular expressions themselves.

Derivatives can be extended to KAT expressions in a very natural
way~\cite{kozen08:kat:coalgebra}: we first define a Boolean function
$\epsilon_\alpha$, that indicates whether an expression accepts the
single atom $\alpha$; this function is then used to define the
derivation function $\delta_{\alpha p}$, that intuitively returns what
remains of the given expression after reading the atom $\alpha$ and
the letter $p$. 
\begin{figure}
  \centering
  \begin{align*}
    \begin{aligned}
      \epsilon_\alpha(x{+}y) &= \epsilon_\alpha(x){+}\epsilon_\alpha(y) \\
      \epsilon_\alpha(x{\cdot}y) &= \epsilon_\alpha(x){\cdot}\epsilon_\alpha(y) \\
      \epsilon_\alpha(x^\star) &= 1\\
      \epsilon_\alpha(q) &= 0 \\
      \epsilon_\alpha(\phi) &=
      \begin{cases}
        1 & \text{if }\alpha\vDash\phi\\
        0 & \text{oth.}
      \end{cases}
    \end{aligned}&&
    \begin{aligned}
      \delta_{\alpha p}(x{+}y) &= \delta_{\alpha p}(x){+}\delta_{\alpha p}(y) \\
      \delta_{\alpha p}(x{\cdot}y) &= 
      \begin{cases}
        \delta_{\alpha p}(x){\cdot}y ~\text{ if }\epsilon_\alpha(x)=0\\
        \delta_{\alpha p}(x){\cdot}y {+} \delta_{\alpha p}(y)
        \text{ oth.}
      \end{cases}\\
      \delta_{\alpha p}(x^\star) &= \delta_{\alpha p}(x)\cdot x^\star\\
      \delta_{\alpha p}(q) &=
      \begin{cases}
        \bddtrue & \text{if }p=q\\
        \bddfalse & \text{oth.}
      \end{cases}\\
      \delta_{\alpha p}(\phi) &= \bddfalse
    \end{aligned}
  \end{align*}
  \vspace{-1em}
  \caption{Explicit derivatives for KAT expressions}
\label{fig:kat:deriv}
\end{figure}
These two functions make it possible to give a coalgebraic
characterisation of the function $G$, we have:
\begin{align*}
  G(x)(\alpha) &= \epsilon_\alpha(x) &
  G(x)(\alpha\,p\,u) &= G(\delta_{\alpha p}(x))(u)\enspace.
\end{align*}

The tuple $\tuple{Reg(\Sigma\uplus B(A)),\delta,\epsilon}$ can be
seen as a deterministic automaton with input alphabet
$At\times\Sigma$, and output set $2^{At}$. Thanks to the above
characterisation, a state $x$ in this automaton accepts precisely the
guarded string language $G(x)$---modulo the isomorphism
$(At\times\Sigma)^\star\to 2^{At}\approx
\pow{(At\times\Sigma)^\star\times At}$.

However, we cannot directly apply the simple algorithm from
Section~\ref{ssec:finite:automata}, because this automaton is not
finite. First, there are infinitely many KAT expressions, so that we
have to restrict to those that are accessible from the expressions we
want to check for equality. This is however not sufficient: we also
have to quotient regular expressions w.r.t.\ a few simple
laws~\cite{kozen08:kat:coalgebra}. This quotient is simple to
implement by normalising expressions; we thus assume that expressions
are normalised in the remainder of this section.

\paragraph{Symbolic derivatives.}
The input alphabet of the above automaton is exponentially large
w.r.t.\ the number of primitive tests:
$At\times\Sigma=2^A\times\Sigma$. Therefore, the simple algorithm
from Section~\ref{ssec:finite:automata} is not tractable in
practice. Instead, we would like to use its symbolic version
(Figure~\ref{alg:symb:equiv}).

The output values (in $(2^{At}={2^A}\to 2)$) are also exponentially
large, and are best represented symbolically, using Boolean BDDs. In
fact, any test appearing in a KAT expression can be pre-compiled into
a Boolean BDD: rather than working with regular expressions over
$\Sigma\uplus B(A)$ we thus move to regular expressions over
$\Sigma\uplus \bdd A 2$, which we call \emph{symbolic KAT
  expressions}. We denote the set of such expressions by $\SKAT$, and
we let $\enc e$ denote the symbolic version of a KAT expression $e$.

Note that there a slight discrepancy here w.r.t.\
Section~\ref{sec:symbolic}: the input alphabet is $2^A\times\Sigma$
rather than just $2^{A'}$ for some $A'$. For the sake of simplicity,
we just assume that $\Sigma$ is actually of the shape $2^{\Sigma'}$;
alternatively, we could work with automata whose transition functions
are represented partly symbolically (for $At$), and partly explicitly
(for $\Sigma$).

\begin{figure}
  \centering
  \begin{align*}
    \begin{aligned}
      \sepsilon(x{+}y) &= \sepsilon(x){\bddlor}\sepsilon(y) \\
      \sepsilon(x{\cdot} y) &= \sepsilon(x){\bddland}\sepsilon(y) \\
      \sepsilon(x^\star) &= 1\\
      \sepsilon(p) &= 0 \\
      \sepsilon(\phi) &= \phi
    \end{aligned}&&
    \begin{aligned}
      \sdelta(x{+}y) &= \sdelta(x)\bddplus\sdelta(y) \\
      \sdelta(x{\cdot} y) &= (\sdelta(x) \bdddot y) \bddplus (\sepsilon(x)\bddtimes\sdelta(y)) \\
      \sdelta(x^\star) &= \sdelta(x) \bdddot x^\star\\
      \sdelta(p) &= \sem{p\mapsto 1, \_\mapsto 0}\\
      \sdelta(\phi) &= 0
    \end{aligned}
  \end{align*}
  \vspace{-1em}
  \caption{Symbolic derivatives for KAT expressions}
\label{fig:kat:symb:deriv}
\end{figure}

We define the symbolic derivation operations in
Figure~\ref{fig:kat:symb:deriv}. 

The output function, $\sepsilon$, has type $\SKAT\to \bdd A 2$, it
maps symbolic KAT expressions to Boolean BDD nodes. The operations
used on the right-hand side of this definition are those on Boolean
BDDs. The function $\sepsilon$ is much more efficient than its
explicit counterpart ($\epsilon$, in Figure~\ref{fig:kat:deriv}): the
set of all accepted atoms is computed at once, symbolically.

The transition function $\sdelta$, has type $\SKAT\to
\bdd{A\uplus\Sigma'}{\SKAT}$. It maps symbolic KAT expressions to BDDs
whose leaves are themselves symbolic KAT expressions. Again, in
contrast to its explicit counterpart, $\sdelta$ computes the all the
transitions of a given expression once and for all. The operations
used on the right-hand side of the definition are the following ones:
\begin{itemize}
\item $n\bddplus m$ is defined by pointwise applying the syntactic sum
  operation from KAT expressions to the two BDDs $n$ and $m$:
  $n\bddplus m=\mes{\phi\mapsto \sem n(\phi)+\sem m(\phi)}$;
\item $n\bdddot x$ syntactically multiplies all leaves of the BDD $n$ by
  the expression $x$, from the right: $n\bdddot x=\mes{\phi\mapsto
    \mes n(\phi)\cdot x}$;
\item $f\bddtimes n$ ``multiplies'' the Boolean BDD $f$ with the BDD
  $n$: $f\bddtimes n=\mes{\phi\mapsto \mes n(\phi) \text{ if } \mes f(\phi)=1,
  0\text{ otherwise}}$.
\item $\mes{q\mapsto 1, \_\mapsto 0}$ is the BDD mapping $q$ to $1$
  and everything else to $0$ ($q\in \Sigma=2^{\Sigma'}$ being casted
  into an element of $2^{A\uplus\Sigma'}$).
\end{itemize}

By two simple inductions, one proves that for all atom $\alpha\in At$,
expression $x\in \SKAT$, and letter $p\in\Sigma$, we have:
\begin{align*}
  \sem{\sepsilon \enc x}(\alpha) &= \epsilon_\alpha(x)\\
  \sem{\sdelta \enc x}(\alpha p) &= \enc{\delta_{\alpha p}(x)}
\end{align*}
(Again, we abuse notation by letting the pair $\alpha p$ denote an
element of $2^{A\uplus\Sigma'}$.) This ensures that the symbolic
deterministic automaton $\tuple{\SKAT,\sdelta,\sepsilon}$ faithfully
represents the previous explicit automaton, and that we can use the
symbolic algorithms from Section~\ref{sec:symbolic}.

\subsection{Partial derivatives}
\label{ssec:partial}

An alternative to Brzozowski's derivatives consists in using
Antimirov' \emph{partial derivatives}~\cite{Antimirov96}, which
generalise to KAT in a straightforward way~\cite{pous:itp13:ra}. The
difference with Brzozowski's derivative is that they produce a
non-deterministic automaton: states are still expressions, but the
derivation function produces a set of expressions. An advantage is
that we do not need to normalise expressions: the set of partial
derivatives reachable from an expression is always finite.

We give directly the symbolic definition, which is very similar to the
previous one:
\begin{align*}
  \spdelta(x{+}y) &= \spdelta(x)\bddcup\spdelta(y) \\
  \spdelta(x{\cdot} y) &= (\spdelta(x) \bddsdot y) \bddcup (\sepsilon(x)\bddstimes\spdelta(y)) \\
  \spdelta(x^\star) &= \spdelta(x) \bddsdot x^\star\\
  \spdelta(p) &=\mes {p\mapsto \set 1, \_\mapsto \emptyset}\\
  \spdelta(\phi) &= \emptyset
\end{align*}
The differences lie in the BDD operations, whose leaves are now sets
of expressions:
\begin{itemize}
\item $n\bddcup m = \mes{\phi\mapsto\sem n(\phi)\cup \sem m(\phi)}$;
\item $n\bddsdot x = \mes{\phi\mapsto\set {x'\cdot x \mid x' \in \sem n(\phi)}}$;
\item $f\bddstimes n = \mes{\phi\mapsto \sem n(\phi) \text{ if } \sem f(\phi)=1,
  \emptyset\text{ otherwise}}$.
\end{itemize}

One can finally relate partial derivatives to Brzozowski's one:
\begin{align*}
  \KA\vdash\Sigma_{x' \in \delta'_{\alpha p}(x)} x' =
  \enc{\delta_{\alpha p}(x)}.
\end{align*}
(We do not have a syntactic equality because partial derivatives
inherently exploit the fact that multiplication distributes over
sums.) Using symbolic determinisation as described in
Section~\ref{ssec:det}, one can thus use the algorithm from
Section~\ref{sec:symbolic} with Antimirov' partial derivatives.

\subsection{Ilie \& Yu's construction}
\label{ssec:ilieyu}

Other automata constructions from the literature can be generalised to
KAT expressions. We can for instance consider Ilie and Yu's
construction~\cite{Ilie-yu-FollowAutomata}, which produces
non-deterministic automata with epsilon transitions with exactly one
initial state, and one accepting state.

We consider a slightly simplified version here, where we elude a few
optimisations and just proceed by induction on the expression. The
four cases are depicted below: $i$ and $f$ are the initial and
accepting states, respectively; in the concatenation and star cases, a
new state $p$ is introduced.

\begin{align*}
  \phi/p:&\quad \xymatrix @R=1em { %
    *+[o][F]{i}\ar[r]_{\phi/p} &
    *+[o][F]{f} }& %
  x\cdot y:&\quad \xymatrix @R=1em { %
    *+[o][F]{i}\ar@/_/@{-}[r]\ar@/^/@{-}[r]_{A(x)} & %
    *+[o][F.]{p}\ar@/_/@{-}[r]\ar@/^/@{-}[r]_{A(y)} & %
    *+[o][F]{f} %
  }\\ %
  x+y:&\quad \xymatrix @R=1em { \\%
    *+[o][F]{i}\ar@/_2.5em/@{-}[r]\ar@/_1em/@{-}[r]_{A(y)} %
               \ar@/^2.5em/@{-}[r]\ar@/^1em/@{-}[r]^{A(x)} & %
    *+[o][F]{f} %
  }& %
  x^\star:&\quad \xymatrix @R=2em @C=2em { &*{}&\\
    *+[o][F]{i}\ar[r]_1 & %
    *+[o][F.]{p}\ar[r]_1 %
    \ar@{-}@(ul,ur)[]^{A(x)} %
    \ar@{-}@(ul,l)[u] %
    \ar@{-}@(ur,r)[u] %
    & *+[o][F]{f} %
  } %
\end{align*}
\smallskip

To adapt this construction to KAT expressions, it suffices to
generalise epsilon transitions to transitions labelled by tests. In
the base case for a test $\phi$, we just add a transition labelled by
$\phi$ between $i$ and $f$; the two epsilon transitions needed for the
star case just become transitions labelled by the constant test $1$.

As expected, when starting from a symbolic KAT expression, those
counterparts to epsilon transitions are labelled by Boolean BDD nodes
rather than by explicit Boolean expressions.

\paragraph{Epsilon cycles.}

The most important optimisation we miss with this simplified
presentation of Ilie and Yu's construction is that we should merge
states that belong to cycles of epsilon transitions. An alternative
to this optimisation consists in normalising first the expressions so
that for all subexpressions of the shape $e^\star$, $e$ does not
contain $1$, i.e., $\sepsilon(e)\neq 1$. Such a normalisation procedure
has been proposed for plain regular expressions by
Br\"uggemann-Klein~\cite{klein93}, it generalises easily to (symbolic)
KAT expressions. For instance, here are typical normalisations:
\begin{align}
  \label{ex1}
  (\phi+p)^\star &\mapsto p^\star \\
  (p^\star+q)^\star &\mapsto (p+q)^\star \\
  \label{ex2}
  ((1+p)(1+q))^\star &\mapsto (p+q)^\star
\end{align}
When working with such normalised expressions, the automata produced
by the above simplified construction have acyclic epsilon transitions,
so that the aforementioned optimisation is unnecessary.
 
According to the example~\eqref{ex1}, it might be tempting to
strengthen example~\eqref{ex2} into $((\phi+p)(\psi+q))^\star \mapsto
(p+q)^\star $. Such a step is invalid, unfortunately. (The second
expression accepts the guarded string $\alpha p \beta$ for all
$\alpha,\beta$, while the starting expression needs
$\beta\vDash\psi$.) This example seems to show that one cannot ensure
that all starred subexpressions are mapped to $0$ by $\sepsilon$. As a
consequence we cannot assume that test-labelled transitions in general
form an acyclic graph.

\subsection{Epsilon transitions removal}
\label{ssec:epsilon}

It remains to eliminate epsilon transitions, so that the powerset
construction can be applied to get a DFA. The usual technique with
plain automata consists in computing the reflexive transitive closure
of epsilon transitions, to precompose the other transitions with the
resulting relation, and to saturate accepting states accordingly.

More formally, let us recall Kozen's matricial representation of
non-deterministic automaton with epsilon
transitions~\cite{kozen94:ka:completeness}, as tuples
$\tuple{n,u,J,N,v}$, where $u$ is a $(1,n)$ 01-matrix denoting the
initial states, $J$ is a $(n,n)$ 01-valued matrix denoting the epsilon
transitions, $N$ is a $(n,n)$ matrix representing the other
transitions (with entries sets of letters in $\Sigma$), and $v$ is a
$(n,1)$ 01-matrix encoding the accepting states.

The language accepted by such an automaton can be represented by
following the matricial product, using Kleene star on matrices:
\begin{align*}
  u\cdot(J+N)^\star\cdot v
\end{align*}
Thanks to the algebraic law $(a+b)^\star=a^\star\cdot (b\cdot
a^\star)^\star$, which is valid in any Kleene algebra, we get
\begin{align*}
  KA\vdash u\cdot(J+N)^\star\cdot v = u\cdot (J^\star N)^\star\cdot (J^\star v)
\end{align*}
We finally check that $\tuple{n,u,0,J^\star N,J^\star v}$
represents a non-deterministic automaton without epsilon transitions.
This is how Kozen validates epsilon elimination for plain automata,
algebraically~\cite{kozen94:ka:completeness}.

The same can be done here for KAT by noticing that tests (or Boolean
BDD nodes) form a Kleene algebra with a degenerate star operation: the
constant-to-1 function. One can thus generalise the above reasoning
to the case where $J$ is a tests-valued matrix rather than a
01-matrix. 

The iteration $J^\star$ of such a matrix can be computed using
standard shortest-path algorithms~\cite{HofnerM12}, on top of the
efficient semiring of Boolean BDD nodes. The resulting automaton has
the expected type:
\begin{itemize}
\item there is a transition labelled by $\alpha p$ between $i$ and $j$
  if there exists a $k$ such that $\alpha \vDash (J^\star)_{i,k}$ and
  $p\in N_{k,j}$. (The corresponding non-deterministic symbolic
  transition function can be computed efficiently using appropriate
  BDD functions.)
\item The output value of a state $i$ is the Boolean BDD node obtained
  by taking the disjunction of all the $(J^\star)_{i,j}$ such that $j$
  is an accepting state (i.e., just $(J^\star)_{(i,f)}$ when using Ilie
    and Yu's construction).
\end{itemize}

\section{Experiments}
\label{sec:exp}

We implemented all presented algorithms, the corresponding library is
available online~\cite{symbolickat:web}.

This allowed us to perform a few experiments and to compare the
various presented algorithms and constructions. We generated random
KAT expressions over two sets of seven primitive tests and seven
atomic elements, with seventy connectives, and excluding the constant
0.  A hundred pairs of random expressions were checked for equality
after being saturated by adding the constant $\Sigma^\star$ (by doing
so, we make sure that the expressions are equivalent, so that the
algorithms have to run their worst case: they cannot stop early thanks
to a trivial counter-example).

Table~\ref{tab:pairs} gives the total number of output tests (e.g.,
line~\ref{line:symb:ce} in Figure~\ref{alg:symb:equiv}) performed by
several combinations of algorithms and automata constructions, as well
as the global running time.

% \begin{table}
%   \centering
%   \begin{tabular}{r|rrr|rrr}
%   & \multicolumn{3}{c}{\code{symb_equiv}} & \multicolumn{3}{c}{\code{dsf_equiv}} \\
%   & Ant. & I.\&Y. & Brz. & Ant. & I.\&Y. & Brz. \\\hline
%   time & 0.5s & 2.7s & 2m06s & 0.5s & 2.7s & 2m05 \\
%   output tests & 1830 & 1867 & 6903 & 1831 & 1867 & 6904 \\
% \end{tabular}
% \caption{Checking expressions w.r.t. their star normal form.}
% \label{tab:ssf}
% \end{table}

\begin{table}
  \centering
  \begin{tabular}{r|rrr|rrr}
  & \multicolumn{3}{c}{\code{symb_equiv}} & \multicolumn{3}{c}{\code{dsf_equiv}} \\
  & Ant. & I.\&Y. & Brz. & Ant. & I.\&Y. & Brz. \\\hline
  time & 1.5s & 7.7s & 2m34 & 1.4s & 7.6s & 1m52 \\
  output tests & 7363 & 7440 & 20167 & 4322 & 4498 & 10255 \\
\end{tabular}
\caption{Checking random saturated pairs of expressions.}
\label{tab:pairs}
\end{table}

One can notice than Antimirov' partial derivatives provide the fastest
algorithms. Ilie and Yu's construction yield approximately the same
number of output tests as Antimirov' partial derivatives, but require
more time, certainly because our implementation of transitive closure
for epsilon removal is sub-optimal. Brzozowski's construction gives
poor results both in terms of time and output tests: the produced
automata are apparently larger, and heavier to compute.

Concerning the equivalence algorithm, one notices that using disjoint
set forests significantly reduces the number of output tests. There is
almost no difference in the timings with the first two constructions,
because most of the time is spent in constructing the automata rather
than checking them for equivalence. This is no longer true with
Brzozowski's construction, for which the automata are sufficiently big
to observe a difference.

\section{Directions for future work}
\label{sec:fw}

Concerning KAT, a natural extension of this work would be to apply the
proposed algorithms to KAT+!B~\cite{GKM14a:KATB} and
NetKAT~\cite{NetKAT14}, two extensions of KAT with important
applications in verification: while programs with mutable tests in the
former case, and network programming in the later case.

KAT+!B has a EXPSPACE-complete equational theory, and its structure
makes explicit algorithms completely useless. Designing symbolic
algorithms for KAT+!B seems challenging.

NetKAT remains PSPACE-complete, and Foster et al. recently proposed a
coalgebraic decision procedure relying on a extension of Brzozowski's
derivatives~\cite{FKMST14a}. To get a practical algorithm, they
represent automata transitions using sparse matrices, which allows for
some form of symbolic treatment. It is important to notice, however,
that by considering (multi-terminal) BDDs here, we go far beyond the
capabilities of sparse transition matrices. Indeed, sparse matrices
just make it possible to factor out those cases where a state has no
successor at all. Consider for instance a KAT expression of the shape
$apx+(\lnot a)py$, where $x$ and $y$ are two non-empty expressions,
possibly using a lot of atomic tests. The derivative of this
expression along a letter $\alpha p$ is either $x$ or $y$ depending on
whether $\alpha(a)$ holds or not. A BDD representation would thus
consist in a single decision node, with two leaves $x$ and $y$. In
contrast, a sparse matrix representation would need to list the
exponentially many atoms together with either $x$ or $y$.

\medskip

Moving away from KAT specificities, we leave open the question of the
complexity of our symbolic variant of Hopcroft and Karp's algorithm
(Figure~\ref{alg:dsf:equiv}). Tarjan proved that Hopcroft and Karp's
algorithm is almost linear in amortised time complexity, and he made
a list of heuristics and path compression schemes that lead to that
complexity~\cite{Tarjan75}. A similar study for the symbolic
counterpart we propose here seems out of reach for now.

% \appendix
% \section{Appendix Title}

% This is the text of the appendix, if you need one.

% \acks

% Acknowledgments, if needed.

%% We recommend abbrvnat bibliography style.

\bibliographystyle{abbrvnat}
\bibliography{short,main}

%% The bibliography should be embedded for final submission.

% \begin{thebibliography}{}
% \softraggedright

% \bibitem[Smith et~al.(2009)Smith, Jones]{smith02}
% P. Q. Smith, and X. Y. Jones. ...reference text...

% \end{thebibliography}

\end{document}